# Amplitude-flattened frequency combs in the moving photonic crystals

## Hang Qu, and Maksim Skorobogatiy*

[1]*Engineering Physics, École Polytechnique de Montréal, Montréal, QC, Canada, H3C 3A7*
*Corresponding author: maksim.skorobogatiy@polymtl.ca*

**In this paper, we explore the spectral properties of the moving hollow-core photonic crystal waveguides (PCWs). We find that frequency combs could be generated in the hollow cores of the moving PCWs. Besides, amplitudes of the sidebands in the generated frequency combs are strongly dependent on the spatial field distribution of the core-guided modes of stationary PCWs. Particularly, we find that introduction of the periodic point defects to the innermost rows of the PCW hollow core would significantly modify the spatial field distributions of the guided modes, leading to appearance of the tightly localized defect modes. This, in turn, leads to the enhancement and equalization of the amplitudes of the side bands in the frequency combs generated by the moving PCWs.**

## 1. Introduction

Photonic crystals (PCs) are the photonic structures in which the dielectric constant exhibits periodic variation in one, two or in all three orthogonal directions [1, 2]. In recent years, the idea of using PCs to alter the dispersion relation of photons has received widespread interest and consideration due to numerous potential applications. It is reported by several groups that passive elements such as slow-light waveguides [3-6], optical delay lines [5, 7, 8], optical buffers [9], etc. can be considerably improved, if constructed on the basis of PCs. Most of the existing research however focused on the stationary PCs, while moving PCs were somewhat overlooked. Due to the relativistic effects, the interaction between light and a moving PC can lead to many intriguing phenomena such as inverse Doppler effect and the generation of frequency combs [10-16]. In our prior studies, we have demonstrated the methodology for analyzing interactions between light and various types of 1D- and 2D- PCs and PCWs [10]. In particular, we found that the monochromatic light launched to the hollow-core of a moving 2D PCW could excite frequency combs. In this paper, we further explore the guiding properties of the moving 2D hollow-core PCWs. We first analyze the core-guided modes of a stationary PCW in the moving reference frame. Then, the mode solutions could be transferred back into the stationary reference frame using Lorentz transformation (LT). We find that amplitudes of the sidebands in the frequency combs generated in the hollow core of moving PCWs strongly depends on the modal distribution of the stationary PCWs. Moreover, we show that the amplitudes of the principal sidebands in the frequency combs could be equalized and enhanced by introducing periodic point defects to the innermost rows of the hollow-core PCWs. Numerical simulations using finite-element software COMSOL confirm that compared to the case of regular moving PCWs (without defects in the core), the moving PCs with the periodic point defects in the core generate frequency combs with the sideband amplitudes 1-2 orders higher and comparable in values. Thus, using moving PCWs with periodic point defects could be a more effective route for the frequency comb generation which also allows enhancement and tuning of the sideband amplitudes. On a general note, using moving PCWs for frequency comb generation allows foregoing of the nonlinear materials and high-power sources, which are currently used for generation of frequency combs. We hope this work would inspire more experimental research relating to the generation of frequency combs and other nonlinear properties of moving PCWs.

**2. Generation of frequency combs in regular moving 2D PCWs and 2D moving PCWs with periodic point defects.**

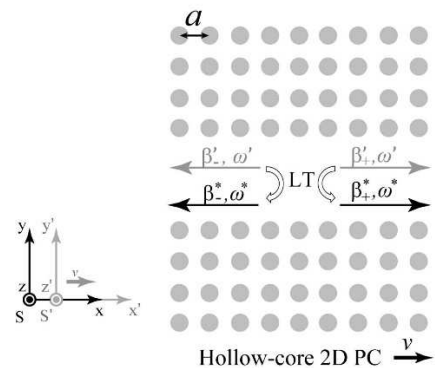

Fig. 1 Schematic of the excitation of guided modes in a moving hollow-core 2D PCW.

We start by analyzing the modal properties of a regular moving 2D PCWs. Consider two reference frames - a stationary frame $S$ with coordinates ($x,y,z,t$) and a moving frame $S'$ with coordinates ($x',y',z',t'$). The corresponding axes of the two frames are mutually parallel, and the frame $S'$ moves at a constant velocity $v$ with respect to the frame $S$ along its $x$-axis. The spacetime in the frame $S'$ could be related to the spacetime in the frame $S$ via Lorentz transformation (LT).

$$t' = \gamma(t - \frac{vx}{c^2})$$
$$x' = \gamma(x - vt),$$
$$y' = y$$
$$z' = z$$
(1)

where $c$ is the velocity of light in the free space, and $\gamma = (\sqrt{1 - v^2/c^2})^{-1}$. In Fig. (1), we present an example of such a waveguide with a hollow core surrounded by a photonic crystal cladding featuring a square lattice of dielectric rods. The waveguide is homogeneous along the $z'$ direction and periodic along the $x'$ and $y'$ directions with the lattice constant $a$. The PC hollow core is a defect introduced into the infinite photonic crystal by removing one row of the dielectric rods. The photonic crystal is moving along the $x$-axis at a constant velocity $v$ in the frame $S$. In the frame $S'$, the photonic crystal is stationary. For the sake of simplicity, we consider TE polarized mode of a waveguide excited by the incident planewave polarized in the $z$ direction. In the case of a finite size (in $x'$ direction) PC, when a planewave is incident onto a photonic crystal waveguide in the frame $S'$, it excites two counter-propagating modes in the waveguide core that have the electric fields in the Bloch form (see page 150 in Ref. [1]):

$$E'_{\pm}(x',y',z',t') = \hat{z}' \cdot U'_{\pm\beta'(\omega')}(x',y')e^{i(\pm\beta'(\omega')x' - \omega't')},$$ (2)

where $\beta'(\omega')$ is a dispersion relation of the core-guided mode of a PC waveguide in the moving reference frame $S'$ in which the waveguide is stationary (see Fig. 2(b)). Also according to the Bloch theorem, $\beta'(\omega')$ is confined to the first Brillouin zone $(-\pi/a, \pi/a]$, and $U'_{\pm\beta'(\omega')}(x',y')$ are the periodic functions along the $x'$-axis with a periodicity of $a$:

$$U'_{\pm\beta'(\omega')}(x'+a, y') = U'_{\pm\beta'(\omega')}(x', y').$$ (3)

Applying LT to the modal form (2), the electric field of the guided mode in the PC hollow core in the frame $S$ could be expressed as (see Eq. (4) in Ref. [10]):

$$E^*(x,y,t) \propto \hat{z} \cdot U'_{\pm\beta'(\omega')}(\gamma(x-vt), y) \cdot e^{i[\pm\beta'(\omega')\cdot\gamma(x-vt) - \omega'\gamma(t - x\frac{v}{c^2})]}$$
$$+ O(v/c)$$
, (4)

where $O(v/c)$ is a correction term on the polynomial order of $v/c$. Since $U'_{\pm\beta'(\omega')}(\gamma(x-vt), y)$ is a periodic function in the $x'$ direction with period $a$, it can be presented in terms of the discrete Fourier series:

$$U'_{\pm\beta'(\omega')}(\gamma(x-vt), y) = \sum_{n=-\infty}^{+\infty} A^n_{\pm\beta'(\omega')}(y)e^{i\frac{2\pi n}{a}\gamma(x-vt)},$$ (5)

where $n$ is an integer, and $A^n_{\pm\beta'(\omega')}$ are the Fourier coefficients. Substituting Eq. (5) into Eq. (4), we obtain the angular frequencies $\omega^*_{\pm,n}$ and the propagation constants $\beta^*_{\pm,n}(\omega^*)$ of the guided modes inside of the moving 2D PCW in the stationary frame $S$:

$$\omega^*_{\pm,n} = \omega'\gamma \pm \beta'(\omega')v\gamma + \frac{2\pi v n\gamma}{a}$$
$$\beta^*_{\pm,n}(\omega^*) = \pm\beta'(\omega')\gamma + \frac{\omega'v\gamma}{c^2} + \frac{2\pi n\gamma}{a}.$$ (6)
$$\omega' = \omega\gamma\left(1 + \frac{v}{c}\right)$$

Here, we assume that the PCW is excited with a monochromatic planewave with frequency $\omega$ in the frame $S$ that is incident from $x = +\infty$. We therefore conclude that two frequency combs are generated inside of the hollow core of a moving 2D PCW. From Eq. (4, 5), we find that the Fourier coefficients $A^n_{\pm\beta'(\omega')}(y)$ represent the amplitudes of the generated sidebands. Also note that these Fourier coefficients could be calculated by applying a Fast Fourier Transform (FFT) to the periodic function $U'_{\pm\beta'(\omega')}(x',y')$. We use the finite element software COMSOL to numerically calculate the electric fields $E'_{\pm}$ of the core-guided mode of the hollow-core PCWs in the moving reference frame $S'$ (see Fig. 2(a)). In our simulation, the lattice constant $a$ of the hollow core PCW is chosen to be 1 μm, and an individual lattice rod is of a radius $R$=0.38 μm and refractive index $n_R$=3. The band diagram of this PCW in the frame $S'$ is presented in Fig. 2(b). As an example, consider photonic modes with $\beta'(\omega') = 0$ (Γ point in Fig. 2(b)). Note that when $\beta'(\omega') = 0$, Eq. (6) could be simplified as:

$$\omega^*_{\pm,n} = \omega'\gamma + \frac{2\pi v n\gamma}{a}$$
$$\beta^*_{\pm,n}(\omega^*) = \frac{\omega'v\gamma}{c^2} + \frac{2\pi n\gamma}{a}.$$ (7)
$$\omega' = \omega\gamma\left(1 + \frac{v}{c}\right)$$

Thus, from Eq. (7), we can find a very simple relation between $\omega^*_{\pm,n}$ and $\beta^*_{\pm,n}(\omega^*)$ as:

$$\beta^*_{\pm,n}(\omega^*) \cdot v = \frac{\omega'v^2\gamma}{c^2} + \frac{2\pi v n\gamma}{a} = -\frac{\omega'}{\gamma} + \omega^*_{\pm,n}.$$ (8)

Applying FFT to $U'_0(x',y')$, we then have the Fourier coefficients $A^n_0(y')$ as shown in Fig. 2(c). One can also calculate the full dispersion relations of the sidebands generated by the core-guided mode of a PCW using Eq. (6), and the results are presented in Fig. 2(d)).

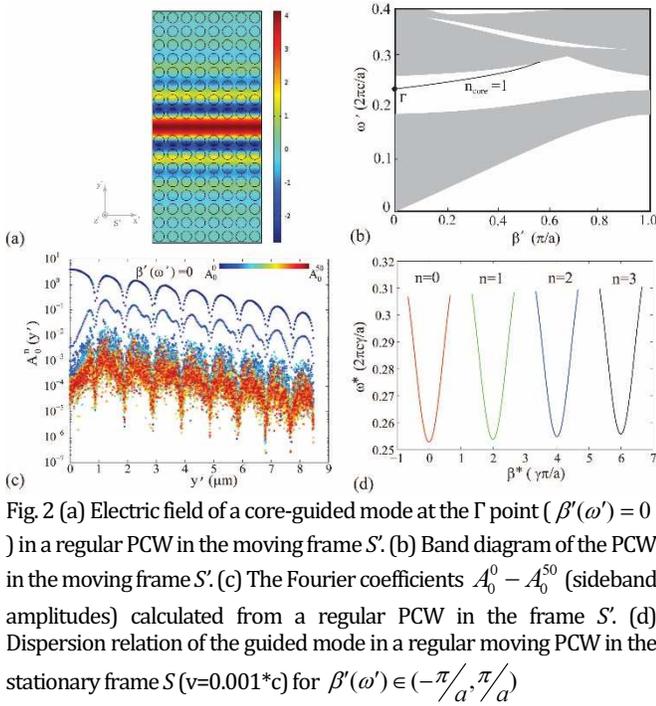

Fig. 2 (a) Electric field of a core-guided mode at the Γ point ($\beta'(\omega') = 0$) in a regular PCW in the moving frame $S'$. (b) Band diagram of the PCW in the moving frame $S'$. (c) The Fourier coefficients $A_0^0 - A_0^{50}$ (sideband amplitudes) calculated from a regular PCW in the frame $S'$. (d) Dispersion relation of the guided mode in a regular moving PCW in the stationary frame $S$ (v=0.001*c) for $\beta'(\omega') \in (-\pi/a, \pi/a)$

From the result shown in Fig. 2(b), we note that the amplitudes of the sidebands in the harmonics are generally three orders smaller than that of the fundamental band. We found that introducing a periodic point defect in the innermost row of the PCW could lead to appearance of the highly localized (in the $x'$ direction) defect states, and thus increase and equalize the amplitudes of the harmonic sidebands. This is easy to understand by noting that sideband amplitudes are simply the Fourier coefficients of the guided mode decomposition in terms of the periodic functions in the $x'$ direction. If a guided mode is strongly localized at a defect site of size $d$, while the PCW period is $a$, then from the basic properties of a Fourier transform, it follows that $\sim 2a/d$ of Fourier coefficients will have comparable values.

As an example, consider a hollow-core PCW featuring a periodic defect in the form of 2 smaller rods in its innermost rows of the PCW core as shown in Fig. 3(a). The defect rod features a smaller radius ($R_d = \sim 0.15a$) and this defect is repeated every $8a$ periods in the $x'$ direction. In the frame $S'$, the electric field $\mathbf{E}'_\pm$ of the core-guided mode in such a waveguide is numerically calculated using COMSOL (Fig. 3(b)). In Fig. 3(c), we also calculate all of the possible states (blue dots) in the PCW; In particular, we show the core-guided 'defect' mode using a solid black curve. Knowing the electric field distribution of the defect mode, we than calculate the corresponding Fourier coefficients $A^n_{\pm\beta'(\omega')}(y)$ using Eq. (5). Compared to the regular moving PCW shown in Fig. 2(a), the moving PCW with a periodic defect point could generate a number of sidebands with much higher and comparable amplitudes. Particularly, we note that the several low-order sidebands have the coefficients (Fig. 3(d)) just one-order smaller than that of the fundamental band, which indicates a significant enhancement in the sideband amplitudes as shown in Fig. 2(b). This is easy to rationalize by noting that modal field of a defect mode has a size $d \sim a$ in the $x'$ direction, while a new period of a PCW in $x'$ direction is $8a$. Thus, we expect $2a/d \approx 16$ sidebands to have comparable amplitudes. Finally, we show the dispersion relation $\beta^*(\omega^*)$ of the core-guided mode of a moving PC waveguide with periodic point defects in the stationary reference frame $S$ (Fig. 3(e)).

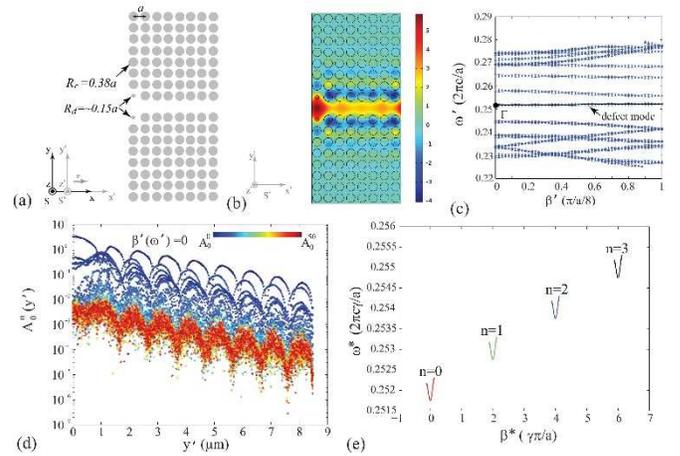

Fig. 3 (a) Schematic of the unit cell of a PCW featuring a periodic point defect with a period $8a$. (b) Electric field in the $S'$ frame of a core-guided mode (at the Γ point, $\beta'(\omega') = 0$) in a PCW featuring a periodic point defect repeated every $8a$. (c) Band diagram of the PCW with a periodic point defect in the moving frame $S'$. Blue dots represent all of the possible states in the PCW, and the solid black curve represents the core-guided "defect" mode. (d) The Fourier coefficients $A_0^0 - A_0^{50}$ (sideband amplitudes) calculated at the Γ point ($\beta'(\omega') = 0$) for a PCW having a periodic point defect. (d) Dispersion relation of the guided mode in a moving PC with a periodic point defect in the stationary frame $S$ (v=0.001*c) for $\beta'(\omega') \in (-\pi/a, \pi/a)$.

Further increasing the spatial separation between the point defects would allow generation of more sidebands with higher and comparable amplitudes. As an example, we simulate the guided mode of a hollow-core PCWs in the moving frame $S'$ that has a periodic point defect with a periodicity of $16a$ (Fig. 4(a)). The corresponding Fourier coefficients $A_0^n(y)$ and dispersion relations of the guided modes in the frame $S'$ are also presented in Fig. 4(b, c). Compared to the moving PC with a point defect having $8a$ periodicity, this moving PC could generate ~32 sidebands that have comparable amplitudes.

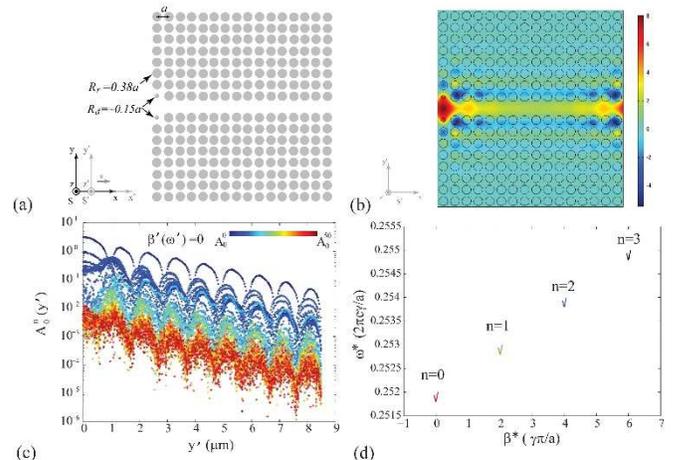

Fig. 4 (a) Schematic of the unit cell of a PCW featuring a periodic point defect with a period of $16a$. (b) Electric field in the $S'$ frame of a guided mode in a moving PC with a periodic point defect of period $16a$. (c) The Fourier coefficients $A_0^0 - A_0^{50}$ calculated at the Γ point ($\beta'(\omega') = 0$) for a PC with a periodic point defect. (d) Dispersion relation of the guided mode in a moving PC with a periodic point defect in the stationary frame $S$ (v=0.001*c) for $\beta'(\omega') \in (-\pi/a, \pi/a)$.

## Conclusions

In summary, in this paper we explore the spectral properties of the moving hollow-core PCWs. We conclude that frequency combs could be generated from the moving hollow-core PCWs. Moreover, the amplitudes of the sidebands in the generated frequency comb are strongly dependent on the field distribution of the core-guided modes of stationary PCWs. Introduction of a periodic point defect to the innermost rows of the PCW hollow core could modify the field distribution of the guided modes of a moving PCW, and lead to the significant enhancement and equalization of the amplitudes of the sidebands in the frequency comb.

## References

1. M. Skorobogatiy, and J. Yang, *Fundamentals of photonic crystal guiding* (Cambridge University Press, 2009).
2. J. D. Joannopoulos, S. G. Johnson, J. N. Winn, and R. D. Meade, *Photonic crystals: molding the flow of light* (Princeton university press, 2011).
3. J. Tang, T. Wang, X. Li, B. Wang, C. Dong, L. Gao, B. Liu, Y. He, and W. Yan, "Wideband and low dispersion slow light in lattice-shifted photonic crystal waveguides," Journal of Lightwave Technology **31**, 3188-3194 (2013).
4. K. Üstün, and H. Kurt, "Ultra slow light achievement in photonic crystals by merging coupled cavities with waveguides," Optics express **18**, 21155-21161 (2010).
5. Y. Wan, K. Fu, C. Li, and M. Yun, "Improving slow light effect in photonic crystal line defect waveguide by using eye-shaped scatterers," Optics Communications **286**, 192-196 (2013).
6. T. Baba, "Slow light in photonic crystals," Nature photonics **2**, 465-473 (2008).
7. Y. A. Vlasov, M. O'boyle, H. F. Hamann, and S. J. McNab, "Active control of slow light on a chip with photonic crystal waveguides," Nature **438**, 65-69 (2005).
8. A. Melloni, A. Canciamilla, C. Ferrari, F. Morichetti, L. O'Faolain, T. Krauss, R. De La Rue, A. Samarelli, and M. Sorel, "Tunable delay lines in silicon photonics: coupled resonators and photonic crystals, a comparison," IEEE Photonics Journal **2**, 181-194 (2010).
9. S. M. Mirjalili, K. Abedi, and S. Mirjalili, "Optical buffer performance enhancement using particle swarm optimization in ring-shape-hole photonic crystal waveguide," Optik-International Journal for Light and Electron Optics **124**, 5989-5993 (2013).
10. H. Qu, Z.-L. Deck-Léger, C. Caloz, and M. Skorobogatiy, "Frequency generation in moving photonic crystals," JOSA B **33**, 1616-1626 (2016).
11. E. J. Reed, M. Soljačić, and J. D. Joannopoulos, "Reversed Doppler effect in photonic crystals," Physical review letters **91**, 133901 (2003).
12. D.-W. Wang, H.-T. Zhou, M.-J. Guo, J.-X. Zhang, J. Evers, and S.-Y. Zhu, "Optical diode made from a moving photonic crystal," Physical review letters **110**, 093901 (2013).
13. J. Chen, Y. Wang, B. Jia, T. Geng, X. Li, L. Feng, W. Qian, B. Liang, X. Zhang, and M. Gu, "Observation of the inverse Doppler effect in negative-index materials at optical frequencies," Nature Photonics **5**, 239-245 (2011).
14. C. Luo, M. Ibanescu, E. J. Reed, S. G. Johnson, and J. Joannopoulos, "Doppler radiation emitted by an oscillating dipole moving inside a photonic band-gap crystal," Physical review letters **96**, 043903 (2006).
15. M. Skorobogatiy, and J. Joannopoulos, "Photon modes in photonic crystals undergoing rigid vibrations and rotations," Physical Review B **61**, 15554 (2000).
16. M. Skorobogatiy, and J. Joannopoulos, "Rigid vibrations of a photonic crystal and induced interband transitions," Physical Review B **61**, 5293 (2000).